\documentclass{elsarticle}
\journal{Physics Letters B}

\usepackage{graphicx,color,amsmath,amssymb}
\usepackage{bm}
\usepackage[colorlinks=true, pdfstartview=FitV, linkcolor=red, citecolor=blue, urlcolor=blue]{hyperref}

\begin{document}

\begin{frontmatter}

\title{Average transverse momentum of hadrons in proton-nucleus collisions in the wounded nucleon model}

\author[ab]{Adam Bzdak}
\address[ab]{RIKEN BNL Research Center, Brookhaven National Laboratory,
Upton, NY 11973, USA}
\ead{abzdak@bnl.gov}

\author[vs]{Vladimir Skokov}
\address[vs]{Department of Physics, Brookhaven National Laboratory, 
Upton, NY 11973, USA}
\ead{vskokov@bnl.gov}

\begin{abstract}
In the wounded nucleon model, a proton-nucleus (p+A) collision is 
a superposition of independent nucleon-nucleon collisions. 
We use this model to calculate the average transverse momentum of pions, 
kaons and protons in high energy p+A collisions. 
For the same number of produced particles, because the number of participants 
differs, in the wounded nucleon
model the transverse momentum of hadrons can differ between p+p and p+A collisions.  
In this model we find that the average transverse momentum in high multiplicity p+A 
collisions depends weakly on the number of produced particles, and underestimates 
the preliminary experimental data at the LHC. The difference is small for pions 
and greater for kaons and protons. The magnitude of this difference is consistent 
with hydrodynamic expectations.
\end{abstract}

\end{frontmatter}

\section{Introduction}

Recent results on the two-particle correlation function in 
p+p~\cite{Khachatryan:2010gv}, p+Pb~\cite{CMS:2012qk,Abelev:2012cya,Aad:2012gla,Chatrchyan:2013nka} 
and Pb+Pb~\cite{Chatrchyan:2011eka,Chatrchyan:2013nka} collisions at the LHC, and in d+Au~\cite{Adare:2013piz} 
collisions at RHIC demonstrated the existence of
the long-range correlations in rapidity. The two-particle correlation function depends on $\cos
(2\Delta \phi )$ and higher order harmonics, where $\Delta \phi =\phi _{1}-\phi _{2}$ is the relative
angle between two particles separated in rapidity. 

An explanation of these phenomena was considered
within the CGC framework for p+p collisions in Ref.~\cite{Dusling:2012cg}, and for
p+Pb collisions in Ref. \cite{Dusling:2012wy}. 
Motivated by 
the convincing success 
in application to A+A collisions, 
hydrodynamics has also been used for a description of p+Pb 
collisions~\cite{Bozek:2011if,BoBro,Bzdak:2013zma,Qin:2013bha}. 
In hydrodynamic models, the dependence of the correlation function on
$\cos (2\Delta \phi )$ and $\cos (3\Delta \phi )$ is naturally present owing to the elliptic and triangular flow. Thus assuming hydrodynamic evolution
in p+Pb collisions, the experimental data can be explained. 
It is also possible that only a joint description involving the CGC framework for the initial 
stage with subsequent hydrodynamic expansion would provide satisfactory agreement 
with the data.

One way to distinguish between different models is to study the average
transverse momentum of hadrons as a function of the number of
produced particles. For example, it is known that hydrodynamic radial flow strongly affects
average transverse momentum, $\langle p_{T} \rangle$, of protons but only slightly that of pions. 
Indeed, the momentum change due to the radial flow can roughly be estimated  
by $m\cdot v_{\mathrm{flow}}$, where $m$ is the mass of the particle and $v_{\mathrm{flow}}$ 
is the flow velocity. 
Thus careful analysis of p+A data can elucidate to what degree hydrodynamics 
plays a role in p+A collisions.

Recently the CMS Collaboration at the LHC revealed its preliminary data on the average transverse 
momentum of pions, kaons and protons as a function of produced particles in p+Pb collisions at 
$\sqrt{s}=5.02$ TeV \cite{talk}. 
Interestingly, the average transverse momentum, $\langle p_{T}\rangle$, 
in p+Pb collisions is quite different in comparison to p+p collisions at similar energies. 
In both systems $\langle p_{T}\rangle$
is a rapidly growing function of the number of produced particles. Moreover, at a given number of particles 
$\langle p_{T}\rangle ^{pA} < \langle p_{T}\rangle ^{pp}$. All this features seem to be consistent 
with the CGC approach, as recently argued in Ref. \cite{McLerran:2013oju}, thus leaving very 
little room for the hydrodynamic evolution.

The data collected at the top RHIC and SPS energies suggest 
that p+A collisions can be viewed, with a good precision, as a superposition of elementary 
p+p collisions. This is strikingly manifested by   
the very good agreement of the p(d)+A data with the prediction of the wounded nucleon 
model \cite{Bialas:1976ed,Bialas:2004su} 
\begin{equation}
N_{ch}^{pA}=N_{\mathrm{part}}\frac{n_{ch}^{pp}}{2},
\label{wnm}
\end{equation}
where $N_{ch}^{pA}$ and $n_{ch}^{pp}$ are the average numbers of produced particles in
p+A and p+p collisions, respectively, and $N_{\mathrm{part}}$ is the number of wounded nucleons.
The wounded nucleon model, and its modifications \cite{ABAB}, proved very successful in 
understanding many features of heavy-ion collisions data.
In this model, the interpretation of Eq. (\ref{wnm}) is straightforward.
In p+A collisions, a nucleon inside a nucleus is struck exactly once and subsequently produces 
a certain number of particles, on the average equal $n_{ch}^{pp} /2$.

In this letter, we apply the wounded nucleon model to investigate how the average transverse momentum 
of particles in p+p collisions reveals itself in p+A collisions.\footnote{The question arises if the 
applicability of the wounded nucleon model to p+A collisions breaks down at the LHC energies, where 
owing to the Lorentz contraction of the colliding nucleus, one could expect formation of coherent 
interacting color fields rather than a superposition of independent elementary nucleon-nucleon collisions~\cite{Larry}.} 
Since $\langle p_{T}\rangle$ in p+p 
collisions is a rapidly changing function of
the number of produced particles \cite{Chatrchyan:2012qb}, it is not clear how $\langle p_{T}\rangle$
in p+A collisions, with many nucleon-nucleon interactions, depends on the number
of produced particles.\footnote{%
Only in the case $\left\langle p_{T}\right\rangle _{N}^{pp}={\rm const}$, the
prediction for p+A  collisions is trivial, that is, $\left\langle p_{T}\right\rangle
_{N}^{pA}={\rm const}$.}
Our calculations may serve as a natural baseline for an onset of {\it{collective}} 
physics (including CGC) irreducible to elementary nucleon-nucleon interactions.  

Our main result is that the average transverse momentum in high multiplicity p+A collisions 
weekly depends on the number of produced particles and underestimates the preliminary
experimental data at the LHC. The difference is rather small for pions ($\sim 100$ MeV) and greater
for kaons ($\sim 300$ MeV) and protons ($\sim 500$ MeV). 
It is interesting that the magnitude of this difference is consistent with
hydrodynamic expectations, where the radial flow pushes
heavier particles to higher momenta.

In the next section, we formulate the problem analytically and present the results based on 
numerical calculations. The last section gives our comments and conclusions. 

\section{Results and discussion}

Suppose  we have $M$ independent sources (wounded nucleons) of particles in p+A
collisions. The single particle distribution for particle species $\alpha$ at a given
transverse momentum, $p_{T}$, is obviously given by the sum of all
sources\footnote{In this letter, we present calculations for positively charged 
pions, kaons, and protons; that is $\alpha=\pi^+, K^+, p$.}
\begin{equation}
f^{pA}_\alpha(p_{T }; N_{1},...,N_{M})= \sum_{i=1} ^{M} f^s_\alpha(p_{T }; N_{i}),
%
\label{fpA}
\end{equation}%
where $f^{pA}_\alpha(p_{T }; N_{1},...,N_{M})$ is the distribution function of particles $\alpha$
in p+A collisions at a given total numbers of particles (number of tracks) $N_i$ coming from each
source $i$.\footnote{Particles $\alpha$ and $N_{i}$ can, in principle, come from different 
rapidity or $p_{T}$ regions.}  
The single-particle distribution function provides 
\begin{equation}
\int f^s_\alpha(p_{T }; {N_{i}}) d^{2}p_{T }=\left\langle N^{\alpha
}\right\rangle _{N_{i}}^{s},\quad \int f^s_\alpha(p_{T }; 
{N_{i}})p_{T }d^{2}p_{T}=\left\langle p_{T}^{\alpha
}\right\rangle _{N_{i}}^{s} \left\langle N^{\alpha
}\right\rangle _{N_{i}}^{s},
\end{equation}%
where $\left\langle N^{\alpha }\right\rangle _{N_{i}}^{s}$ is the average
number of particles $\alpha$ from the $i$-th source  at a given total number of particles, $N_{i}$, from this source; 
$\left\langle p_{T}^{\alpha
}\right\rangle _{N_{i}}^{s}$ is the average transverse momentum. 

Using Eq.~\eqref{fpA}, 
the average transverse momentum in p+A collisions can be readily derived
\begin{equation}
\left\langle p_{T}^{\alpha }\right\rangle _{N_{1},...,N_{M}}^{pA}=\frac{1}{\langle N^\alpha\rangle^{pA}_{N_1,\dots,N_M}}%
\sum_{i=1}^M \left\langle p_{T}^{\alpha }\right\rangle _{N_{i}}^{s}\left\langle N^{\alpha
}\right\rangle _{N_{i}}^{s}, 
\label{ptpA}
\end{equation} 
where we defined the total number of particles $\alpha$ (from all sources) by   
\begin{equation}
\langle N^\alpha\rangle^{pA}_{N_1,\dots,N_M} = \sum_{i=1}^M \left\langle N^{\alpha
}\right\rangle _{N_{i}}^{s}.  
\label{NpA}
\end{equation}
It is natural to expect that mean number of particles $\alpha$, 
$\left\langle N^{\alpha
}\right\rangle _{N_{i}}^{s}$, from one source 
is proportional to the total number of particles (or tracks) from a given source $N_i$, that is  
\begin{equation}
\left\langle N^{\alpha }\right\rangle _{N_{i}}^{s}\propto N_{i}.
\label{total_to_alpha}
\end{equation}
This simplifies Eq.~\eqref{ptpA} into%
\begin{equation}
\left\langle p_{T}^{\alpha }\right\rangle _{N_{1},...,N_{M}}^{pA}={%
 \sum_{i=1}^M \left\langle p_{T}^{\alpha}\right\rangle _{N_{i}}^{s} N_{i}}\left/{ \sum_{i=1}^M N_{i}}\right. .
\end{equation}

The average transverse momentum  in p+A collisions at a given total number
of produced particles $N_{\rm tot}$ can be obtained by sampling distributions of $N_i$ such
that $N_{\rm tot}=\sum_i N_i$. 
In other words, 
\begin{eqnarray}
&&\left\langle p_{T}^{\alpha }\right\rangle _{N_{\mathrm{tot}}}^{pA}=%
\\&&
\frac{1}{\cal N}
\sum_{M=1}^{A+1} P_G(M) 
\sum_{N_{1},...,N_{M}} 
\frac{ \sum_{i=1}^{M} 
\left\langle p_{T}^{\alpha }\right\rangle _{N_{i}}^{s}\left\langle N^{\alpha }\right\rangle
_{N_{i}}^{s}}
  {   \sum_{i=1}^M   \left\langle N^{\alpha }\right\rangle _{N_{i}}^{s}} \prod_{i=1}^M P(N_{i})\delta {\left(\sum_{i=1}^M N_{i}-N_{\mathrm{tot}} \right)}, \nonumber
\end{eqnarray}
where $P_G(M)$ is the distribution of the number of participants in p+A
collisions and ${\cal N}$ is the proper normalization given by 
\begin{equation}
{\cal N} = \sum_{M=1}^{A+1} P_G(M) 
\sum_{N_{1},...,N_{M}} 
\prod_{i=1}^M P(N_{i})\delta {\left(\sum_{i=1}^M N_{i}-N_{\mathrm{tot}} \right)}. 
\end{equation}
In this paper, $P_G(M)$ is defined by the Glauber Monte Carlo simulations. 
The probability $P(N_{i})$ is the multiplicity distribution of produced
particles from the $i$-th source. Using Eq.~\eqref{total_to_alpha} we obtain
\begin{eqnarray}
&&\left\langle p_{T}^{\alpha }\right\rangle _{N_{\mathrm{tot}}}^{pA}=%
\\&& \nonumber
\frac{1}{\cal N}
\sum_{M=1}^{A+1}P_G(M)\sum_{N_{1},...,N_{M}}
\frac{   \sum_{i=1}^M  
\left\langle p_{T}^{\alpha }\right\rangle _{N_{i}}^{s}  N_{i}
 } {   \sum_{i=1}^M   N_{i}} \prod_{i=1}^M P(N_{i})\delta {\left(\sum_{i=1}^M N_{i}-N_{\mathrm{tot}} \right)}.
 \label{ptpAF}
\end{eqnarray}

To calculate 
$\left\langle p_{t}^{\alpha }\right\rangle _{N_{\mathrm{tot}}}^{pA}$ we apply 
the Glauber Monte Carlo model. As usual we first sample positions 
of nucleons in a nucleus 
according to the Woods Saxon distribution with parameters given in Ref.~\cite{Alver:2008aq}. 
The number of participants 
is then computed taking inelastic cross-section for p+p collisions to 
be $\sigma=65$ mb~\cite{Zsigmond:2012vc}.\footnote{We checked that hard-sphere and Gaussian inelastic 
profiles \cite{ABAB,Rybczynski:2011wv} lead to the same results.}
To each wounded nucleon we prescribed the number of produced particles, in $|\eta|<2.4$, sampled according to the 
negative binomial distribution\footnote{Average number of particles and $k$ parameter in the 
negative binomial distribution equal to half of 
the values measured in p+p collisions.} with parameters obtained from experimental 
data \cite{Khachatryan:2010nk}, analysed in Ref.~\cite{Ghosh:2012xh}.
 
In the case of p+p collisions we have 
\begin{equation}
\left\langle p_{T}^{\alpha }\right\rangle _{N_{1},N_{2}}^{pp}=\frac{%
\left\langle p_{T}^{\alpha }\right\rangle _{N_{1}}^{s}N_{1}+\left\langle
p_{T}^{\alpha }\right\rangle _{N_{2}}^{s}N_{2}}{N_{1}+N_{2}}.
\end{equation}
If $N_{1}=N_{2}$ we obtain $\left\langle p_{T}^{\alpha }\right\rangle
_{N_{1},N_{1}}^{pp}=\left\langle p_{T}^{\alpha }\right\rangle _{N_{1}}^{s}$, if 
$N_{1}\gg N_{2}$ we get $\left\langle p_{T}^{\alpha }\right\rangle
_{N_{1},N_{2}}^{pp} \approx \left\langle p_{T}^{\alpha }\right\rangle _{N_{1}}^{s}$. In
the former case $\left\langle p_{T}^{\alpha }\right\rangle _{N_{\mathrm{tot}%
}=2N_{1}}^{pp}=\left\langle p_{T}^{\alpha }\right\rangle _{N_{1}}^{s}$ and the
latter $\left\langle p_{T}^{\alpha }\right\rangle _{N_{\mathrm{tot}%
} \approx N_{1}}^{pp} \approx \left\langle p_{T}^{\alpha }\right\rangle _{N_{1}}^{s}.$ We
checked that for all practical purposes%
\begin{equation}
\left\langle p_{T}^{\alpha }\right\rangle _{N_{i}}^{s} \approx \left\langle p_{T}^{\alpha
}\right\rangle _{N_{\mathrm{tot}}=\kappa N_{i}}^{pp}
\label{app}
\end{equation}
with $\kappa=1.5$.
We use this approximation in our Monte Carlo calculations of p+A collisions. 
The input for the average transverse momentum~\eqref{app} in p+p collisions is taken from
the CMS data published in Ref.~\cite{Chatrchyan:2012qb}.

\begin{figure}
\centerline{\includegraphics[width=0.52\textwidth]{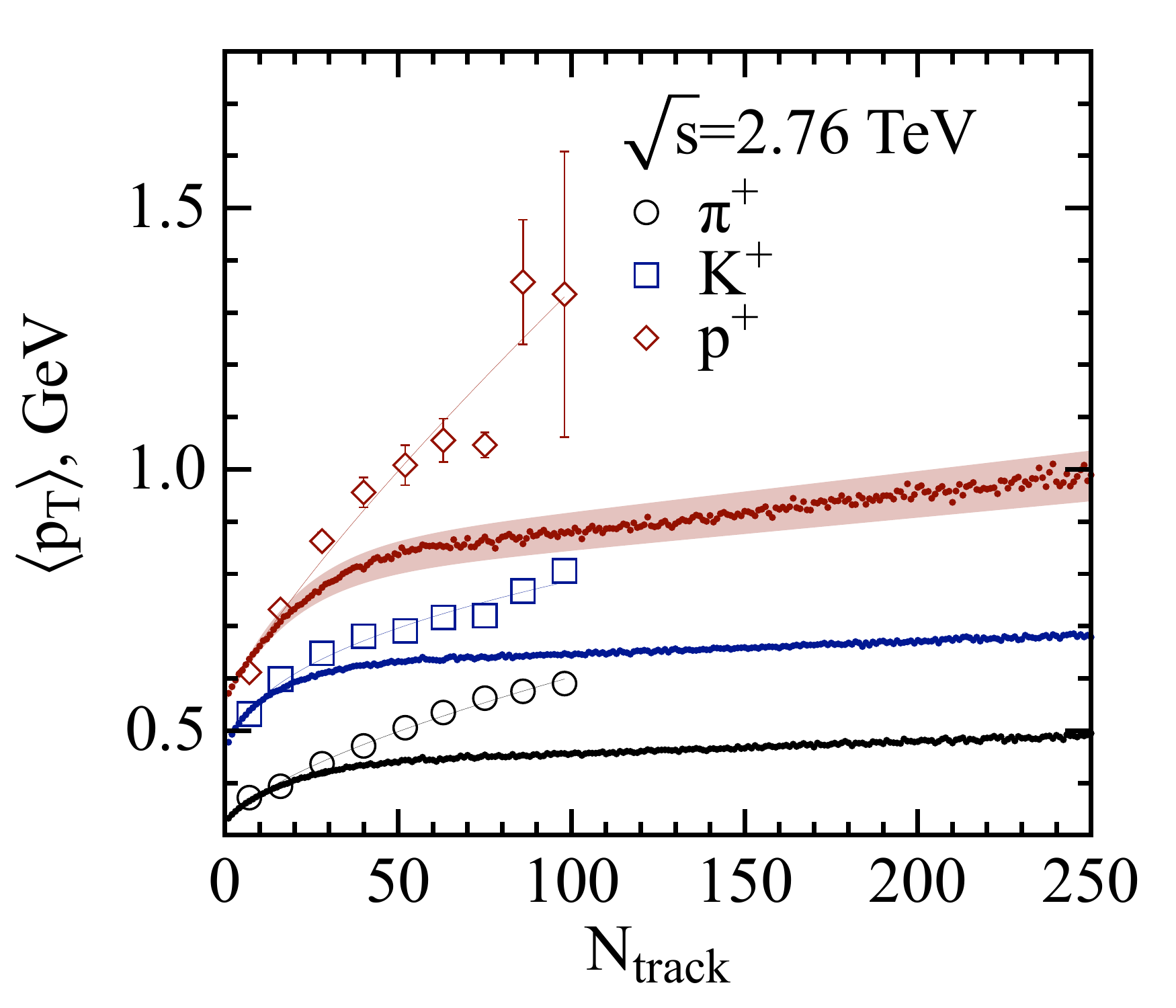} 
\includegraphics[width=0.52\textwidth]{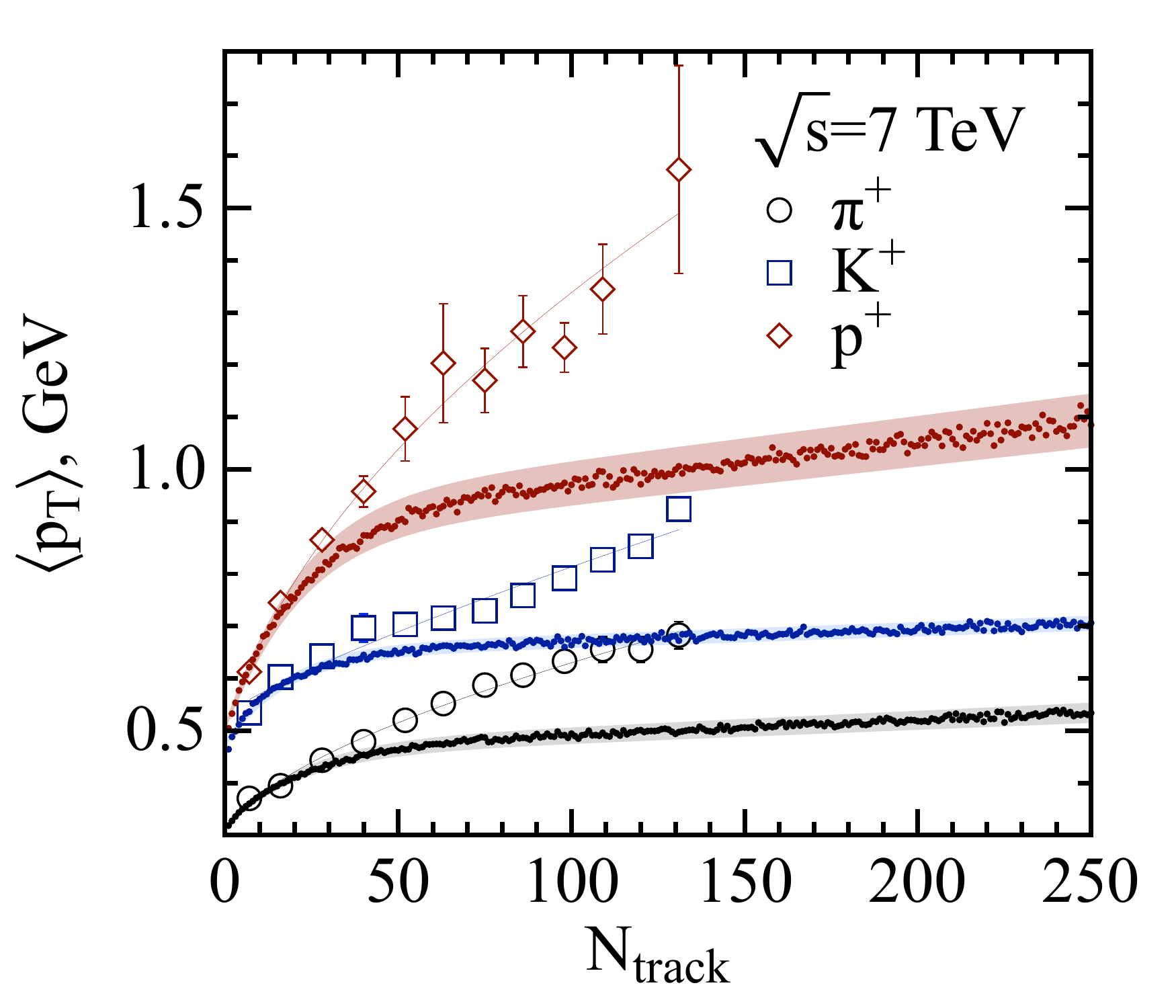}}
\caption{
The average transverse momentum as a function of $N_{\rm track}$, measured in $|\eta|<2.4$, 
for different particle species in 
the wounded nucleon model for p+Pb collisions (small filled points) at $\sqrt{s}=2.76$ TeV and $7$ TeV.  
The band shows theoretical uncertainty of the model due to variation of the coefficient 
$\kappa$ in Eq.~\eqref{app} within  the range $[1.3, 1.7]$.
The large open symbols are the experimental data on average transverse momentum in p+p collisions, 
used as  an input in our calculations.
}
\end{figure}

In Figure 1, we show the results for p+Pb collisions obtained in the wounded nucleon model 
(small filled points) in comparison to p+p experimental data (open symbols), which were used 
as an input in our calculations. 
We carried out the computations for two energies $\sqrt{s}=2.76$ TeV and  $\sqrt{s}=7$ TeV, for 
positively charged pions, kaons and protons. Figure 1 shows that p+Pb
calculations are always below experimental p+p data points. This can be understood in the following  simple 
way. In an ordinary p+A collision the number of participants and consequently number 
of produced particles is larger than in an ordinary p+p collision. Thus, comparing both systems 
at the same number of produced particles, say $N_{\rm track}=100$, a typical nucleon-nucleon interaction 
in a p+Pb collision produces less than $100$ particles. Consequently, owing to the non-flat 
dependence of average transverse momentum in p+p as a function of $N_{\rm track}$, our p+Pb results 
are below the p+p data points. It is also worth noticing that our p+Pb results are significantly 
flatter then the experimental p+p data.

The CMS Collaboration has collected preliminary data in p+Pb collisions at $\sqrt{s}=5.02$ TeV~\cite{talk}. 
Since our model results do not change 
strongly in the energy range from 2.76 TeV to 7 TeV, we will compare the preliminary p+Pb 
experimental data with our model at $\sqrt{s}=$ 7 TeV. 
As shown in Figure 2, our model results (bands) for protons, kaons and pions, are below the 
preliminary data (full symbols).
For example, for $N_{\rm track}=200$ our model calculations have a difference with 
experimental data of  about 100 MeV for pions, 300 MeV for kaons and 500 MeV for protons.
First, this suggests that high multiplicity p+Pb collision is irreducible 
to a superposition of p+p collisions and that collective/coherent effects must be accounted for. 
Second, the apparent discrepancy 
between the preliminary experimental data and our results may be explained by some 
mechanism which shifts particles to higher momentum proportionally to their masses, as one expects 
in the hydrodynamic models, where, the shift in the transverse momentum is roughly proportional 
to $m\cdot v_{\mathrm{flow}}$.\footnote{As seen in the mean transverse momentum difference between peripheral and central 
collisions in nucleus-nucleus collisions, see e.g., Ref.~\cite{Abelev:2013vea}.} The final 
conclusion, however, requires quantitative analysis of the data in a 
hydrodynamic model. 

\begin{figure}
\centerline{\includegraphics[width=0.52\textwidth]{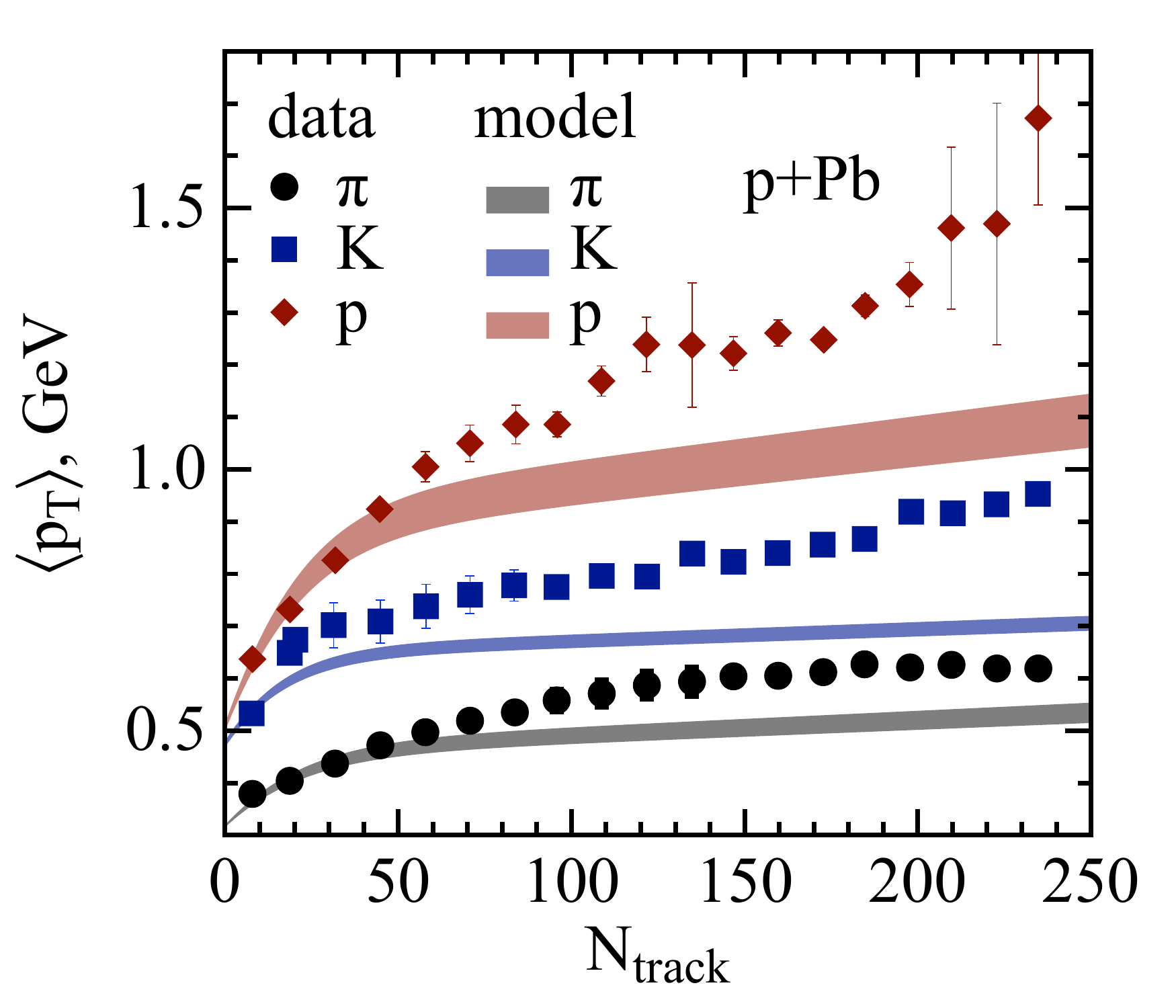}} 
\caption{
The average transverse momentum as a function of $N_{\rm track}$ for protons, kaons and pions  
in the wounded nucleon model for p+Pb collisions  at $\sqrt{s}=7$ TeV compared to the  
preliminary experimental data by the CMS Collaboration at $\sqrt{s}=5.02$ TeV. 
}
\end{figure}
\section{Comments and conclusions}
Using the wounded nucleon model, based on the available experimental data of p+p collisions, we 
calculated the average transverse momentum of positively charged pions, kaons and protons 
in p+Pb collisions. 
The model results show that at a given number of particles (tracks) the 
average transverse momentum in p+Pb collisions is significantly smaller then the 
one in p+p collisions. 
We also compared our results with the preliminary CMS data in p+Pb collisions and we observed 
that the wounded nucleon model underestimates the data. This demonstrates that in high multiplicity p+Pb 
collisions at the LHC energy we encounter physics which cannot be reduced to a superposition of 
elementary p+p collisions. The magnitude of the difference between our model results and the preliminary 
CMS data is consistent with hydrodynamic expectations, and is small for pions and greater for 
kaons and protons.

Finally we comment that similar studies
in central d+Au collisions would be very informative.\footnote{We thank
Anne Sickles for
pointing this out to us.} In this case, Eq. (\ref{wnm}) is experimentally verified to 
a good precision
\cite{Bialas:2004su}, that is, a d+Au collision at $\sqrt{s}=200$ GeV can be
viewed as a superposition of elementary nucleon-nucleon collisions, when the
number of particles is considered. The possible hydrodynamic effects are not
expected to change the number of particles. 
Thus, it would be useful to
analyze the measured PHENIX data~\cite{Adler:2006xd,Adare:2013esx} and
extract the average transverse momentum in p+p and d+Au collisions at
different centralities. If there is no radial flow, we expect to
observe%
\begin{equation}
\left\langle p_{T}^{\alpha }\right\rangle ^{dAu}=\left\langle
p_{T}^{\alpha}\right\rangle ^{pp}
\end{equation}%
for all centralities and all particles species $\alpha $. Here $\left\langle
p_{T}^{\alpha }\right\rangle ^{pp}$ is averaged over all numbers of produced
particles (minimum bias) and
$\left\langle p_{T}^{\alpha }\right\rangle ^{dAu}$ is averaged over
all numbers of
produced particles within a given
centrality class. A possible radial flow in d+Au collisions would increase
$\left\langle p_{T}^{\alpha }\right\rangle ^{dAu}$ so that
\begin{equation}
\left\langle p_{T}^{\alpha }\right\rangle ^{dAu}-\left\langle p_{T}^{\alpha
}\right\rangle ^{pp}\approx m_{\alpha }v_{\rm flow}
\end{equation}
is small for pions and larger for kaons and protons.

\section*{Acknowledgments}
We thank Larry McLerran, Robert Pisarski and Raju Venugopalan for stimulating 
discussions and constructive criticism. 
A.B. is supported through the RIKEN-BNL Research Center. V.S. is supported
by the U.S. Department of Energy under contract \#DE-AC02-98CH10886.

\end{document}